\documentclass[
 preprint, amsmath,amssymb,
 aps,physrev,
]{revtex4-2}

\usepackage{graphicx}
\usepackage{subcaption}
\usepackage{algorithm}
\usepackage{amsmath}
\usepackage{placeins}
\usepackage[colorlinks=true, linkcolor=blue, citecolor=blue, urlcolor=blue]{hyperref}

\usepackage{xcolor}
\usepackage[utf8]{inputenc} 

\begin{document}

\preprint{APS/123-QED}

\title{\textbf{Coherent control of optomechanical entanglement and steering via dual parametric amplification} }%
\author{Jinhao Jia}
\author{Yingru Li}
\author{Ran Liang}
\author{Mei Zhang}
\email{zhangmei@bnu.edu.cn}
\affiliation{School of Physics and Astronomy, Beijing Normal University,
Beijing 100875, China}
\affiliation{Applied Optics Beijing Area Major Laboratory,
Beijing Normal University, Beijing 100875, China}
\affiliation{Key Laboratory of Multiscale Spin Physics (Ministry of Education),
Beijing Normal University, Beijing 100875, China}

\date{\today}

\begin{abstract}
We propose a coherent control scheme for engineering quantum correlations in a cavity optomechanical (COM) system consisting of a driven optical cavity with an embedded nonlinear medium and a membrane, assisted by a coherent feedback loop. The nonlinear medium and the membrane are pumped to implement optical and mechanical parametric amplifications with controllable modulation frequencies and pump amplitudes. Within the stable regime identified by Floquet analysis, the feedback-induced enhancement of optomechanical entanglement persists when the period-averaged mean-field intracavity photon number is matched, showing that the enhancement cannot be attributed simply to an increase in this quantity.
A systematic comparison of the four configurations further reveals a nonadditive interplay between dual parametric amplification and coherent feedback. Varying the amplitude reflectivity of the beam splitter and the feedback loop phase enables effective control of optomechanical entanglement and directional EPR steering, including a transition from one-way to two-way steering. The combined scheme also improves the robustness of both correlations against thermal noise, providing a potential route toward robust quantum-state engineering in COM systems.
\end{abstract}

\maketitle

\section{\label{sec:level1}Introduction}

Quantum entanglement \cite{Laure2021,Shlomi2021,Tao2020,lv2018,Horo2009}, as one of the most fundamental features of quantum mechanics, is a key source for quantum metrology \cite{Pezz2018,zhang2015}, quantum computation \cite{Graham2022,Bartolucci2023}, and quantum communication \cite{Ursin2007,Piveteau2022}. It has been demonstrated in atoms \cite{Ritter2012}, ions \cite{Jost2009}, and cavity optomechanical (COM) systems \cite{Farace2012,yang2024,peng2023,Li2017}, where the interaction between an optical cavity and mechanical motion is mediated by radiation pressure \cite{Genes2008,Aspelmeyer2014}. COM systems have enabled a variety of important phenomena and applications, such as mechanical squeezing \cite{Szor2011,lv2015,Bai2020,Han2019}, optomechanically induced transparency \cite{Yan2020,Lai2020}, and mechanical resonator cooling \cite{Zoepfl2023,Chang2024}. Beyond entanglement, Einstein-Podolsky-Rosen (EPR) steering is a form of quantum nonlocality intermediate between entanglement and Bell nonlocality \cite{Reid2009,Caval2017,Guan2024,Bai2024}. Several schemes to generate steady-state one-way steering in COM systems have also been explored \cite{Tan2017,Lijie2017}. To protect and enhance the fragile quantum correlations in COM platforms, substantial efforts have been devoted to developing various approaches. Among these, parametric driving enabled by optical parametric amplification (OPA) and mechanical parametric amplification (MPA) provides a versatile and effective means to enhance nonlinearity and access rich classical and quantum dynamics \cite{Zhu2025,Shao2024,Xiao2024,Hu2020,Wei2024,Liu2025}.

Feedback control is also a viable strategy for effectively enhancing quantum correlations in COM systems by extracting and re-injecting quantum information \cite{Peng2015,Luo2020,Harwod2021,Ebrahimi2022,Wei2021}. Broadly, feedback mechanisms can be divided into two categories: measurement-based feedback \cite{Rossi2017,Qiu2022} and coherent feedback \cite{Lloyd2000,Schmid2022}. The former is typically complicated, as it relies on measurement and subsequent classical processing, which can introduce additional noise and backaction. In contrast, coherent feedback avoids explicit measurement and does not introduce measurement-induced noise. It has therefore attracted interest for applications such as improving resonator cooling \cite{Frimmer2016}, protecting qubit coherence \cite{Hirose2016}, and enhancing entanglement and steering \cite{Li2017,peng2023}.

Dual parametric amplification \cite{yang2024} and coherent feedback \cite{Li2017,peng2023} have each provided effective approaches to engineering quantum correlations in COM systems. Building on these advances, we investigate their combined action in a periodically driven system.
In particular, the multi-field-driven scheme of Ref.~\cite{yang2024} employs OPA and MPA to enhance and control optomechanical entanglement, but does not include coherent reinjection, which simultaneously modifies the effective cavity drive, decay rate, detuning, and coupling to optical input noise. It therefore remains important to determine whether the feedback-induced enhancement persists at the same period-averaged mean-field intracavity photon number and whether coherent feedback and dual parametric amplification produce a nonadditive response within the dynamically stable regime.

To address these issues, we consider a Fabry–Pérot (FP) cavity that contains a nonlinear $\chi^{(2)}$ medium and a membrane and is coupled to a coherent feedback loop. The loop consists of a highly reflecting mirror (HRM) and a controllable beam splitter (CBS), which together coherently reinject the cavity output field into the cavity. The nonlinear medium is optically pumped to implement OPA, whereas the membrane is parametrically driven to implement MPA. An external laser drives the cavity through the beam splitter. The amplitude reflectivity $r_b$ of the beam splitter and the feedback phase $\theta$ control the reinjected field and thereby modify the effective cavity parameters and the coupling to optical input noise. For the commensurate parametric-pump frequencies considered here, the linearized dynamics are periodic. We therefore determine dynamical stability from the Floquet multipliers and evaluate the quantum correlations using the asymptotic periodic covariance matrix.

Within the identified stable regime, we first investigate optomechanical entanglement as a function of the OPA–MPA modulation-frequency ratio, modulation-amplitude ratio, and CBS reflectivity. A comparison of the configurations with and without coherent feedback shows that the feedback-induced enhancement of entanglement persists when the period-averaged mean-field intracavity photon number is matched and therefore cannot be attributed simply to an increase in this quantity. The covariance-matrix invariants further characterize the accompanying changes in the Gaussian state. A systematic comparison of the reference configuration, dual parametric amplification alone, coherent feedback alone, and the combined scheme further reveals a nonadditive interplay between dual parametric amplification and coherent feedback: for the representative parameters considered here, coherent feedback alone suppresses entanglement, whereas it enhances entanglement in the presence of both parametric amplifications. Varying the beam splitter reflectivity and feedback phase provides additional control over optomechanical entanglement and directional EPR steering, including a transition from one-way to two-way steering. Finally, the temperature-dependent analysis shows that the combined scheme improves the robustness of both entanglement and EPR steering against thermal noise compared with dual parametric amplification alone. These results may provide a useful route toward the robust generation and manipulation of quantum correlations in COM systems, with potential applications in continuous-variable quantum information processing.

\section{\label{sec:level2}System and dynamics}

\begin{figure}[htbp]
  \begin{subfigure}[b]{0.5\linewidth}
    \centering
    \includegraphics[width=\linewidth]{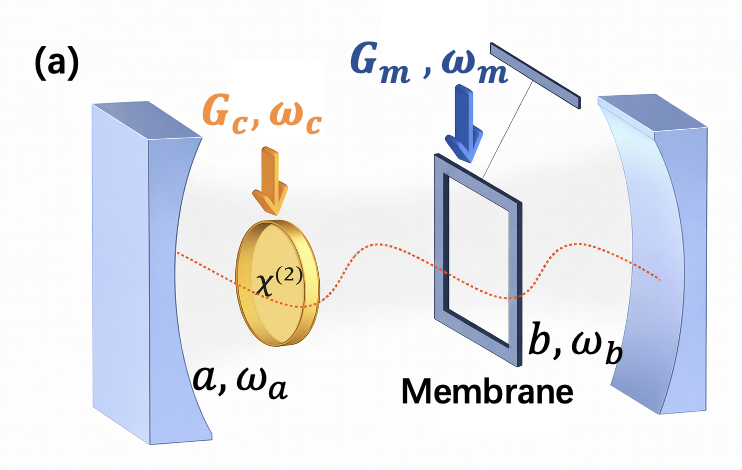} 
 
  \end{subfigure}
  \hfill
  \begin{subfigure}[b]{0.5\linewidth}
    \centering
    \includegraphics[width=\linewidth]{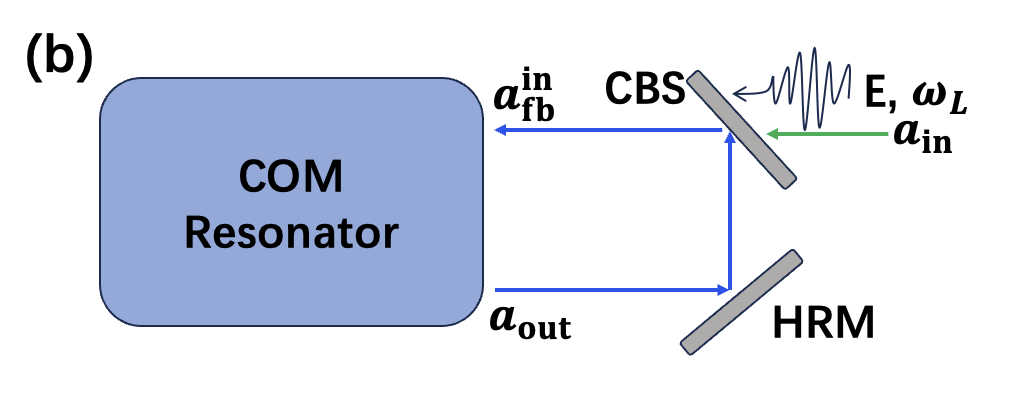}
    
  \end{subfigure}
  \caption{Schematic diagram of the proposed COM system with the embedded membrane and the $\chi^{(2)}$ nonlinear medium, coupled to the coherent feedback loop.
(a) The nonlinear medium is driven by a coherent field with amplitude $G_c$ and frequency $\omega_c$ to implement OPA, while the membrane is subjected to a periodic pump with amplitude $G_m$ and frequency $\omega_m$ to implement MPA.
(b) The coherent feedback loop consists of the HRM and the CBS. The external laser field with amplitude $E$ drives the cavity via the CBS. Here $ a_{\mathrm{in}}$ denotes the optical input vacuum noise associated with the zero-point fluctuation entering the cavity through the CBS. The blue arrows indicate that the optical output field $a_{\mathrm{out}}$ reflected by the HRM is fed back into the cavity through the CBS.}\label{fig1}
  \end{figure}  

As shown in Fig.~\ref{fig1}, we consider a COM setup in which a nonlinear $\chi^{(2)}$ medium and a membrane are embedded inside an FP cavity with a feedback loop. The FP cavity supports a photon mode with frequency $\omega_a$. The membrane acts as a mechanical oscillator supports a phonon mode with effective mass $m$ and frequency $\omega_b$ \cite{peng2023}. The cavity is pumped by an external laser field with amplitude $E$, transmitted through a CBS with amplitude transmissivity $t_b$ and amplitude reflectivity $r_b$, satisfying $t_b^2+r_b^2=1$ \cite{Li2017}. $E=\sqrt{2\kappa_{a}P_{l}/(\hbar\omega_{l})}$ is the amplitude of the laser, where $\kappa_{a}$ is the decay rate of the optical cavity, $P_{l}$ ($\omega_{l}$) is the laser power (frequency) \cite{Liu2017}. A coherent field with amplitude $G_c$, frequency $\omega_c$, and phase $\theta_c$ is applied to the nonlinear medium inside the cavity, giving rise to OPA \cite{Zhu2025,Shao2024}. The membrane is simultaneously subjected to a periodic pump of frequency $\omega_m$, amplitude $G_m$, and phase $\theta_m$, which periodically modulates the spring constant and produces MPA \cite{Liu2025,yang2024}. The Hamiltonian of the COM system in a frame rotating with respect to the
driving laser frequency $\omega_l$ reads \cite{Zhu2025}
\begin{align}
H/\hbar &=
\Delta_a a^\dagger a
+\omega_b b^\dagger b
-g a^\dagger a(b^\dagger+b)
+\mathrm{i}G_c\left(
a^{\dagger 2}e^{-\mathrm{i}\Delta_c t+\mathrm{i}\theta_c}
-a^2e^{\mathrm{i}\Delta_c t-\mathrm{i}\theta_c}
\right)
\nonumber\\
&\quad
+\mathrm{i}G_m\left(
b^{\dagger 2}e^{-\mathrm{i}\omega_m t+\mathrm{i}\theta_m}
-b^2e^{\mathrm{i}\omega_m t-\mathrm{i}\theta_m}
\right)
+
\mathrm{i}\left(
E_{\mathrm{fb}}a^\dagger
-E_{\mathrm{fb}}^{*}a
\right) ,
\end{align}
where $\Delta_a=\omega_a-\omega_l$ and $\Delta_c=\omega_c-2\omega_l$. Here, $a (a^{\dagger})$ and $b (b^\dagger)$ denote the annihilation (creation) operators of the cavity photon and mechanical phonon modes, respectively. They obey the bosonic commutation relations $[a,a^\dagger]=[b,b^\dagger]=1$. The parameter $g$ denotes the single-photon optomechanical coupling strength.
$E_{\mathrm{fb}}=t_bE/(1+r_be^{\mathrm{i}\theta})$ denotes the effective coherent driving
amplitude in the presence of the feedback loop.

In the feedback channel, the output field is reflected by the HRM and then fed back into the optical cavity through the beam splitter. By using the standard input-output relation, the output field is given by $a^{\mathrm{out}}=\sqrt{2\kappa_{a}}a-\tilde{a}_{\mathrm{fb}}^{\mathrm{in}}$ \cite{Agarwal2013}. The feedback-modified input field can be described as a superposition of the original input noise $a^{\mathrm{in}}$ and the returned output field. These two contributions are coherently mixed by the CBS before re-entering the cavity, so that the input-field operator can be written as $\tilde{a}_{\mathrm{fb}}^{\mathrm{in}}=r_be^{\mathrm{i}\theta}a^{\mathrm{out}}+t_b a^{\mathrm{in}}$ \cite{Harwod2021,Wei2021}. The parameter $\theta$ denotes the total feedback-loop phase at the driving laser frequency. Combining the above two input-output relations, we obtain
\begin{equation}
\tilde{a}_{\mathrm{fb}}^{\mathrm{in}}
=
\frac{\sqrt{2\kappa_a}r_be^{\mathrm{i}\theta}}
{1+r_be^{\mathrm{i}\theta}}a
+
\frac{t_b}
{1+r_be^{\mathrm{i}\theta}}a^{\mathrm{in}}.
\end{equation}
The effective optical input noise is given by $a_{\mathrm{fb}}^{\mathrm{in}}
=t_b/(1+r_be^{\mathrm{i}\theta})a^{\mathrm{in}}$.
Substituting the self-consistent feedback field into the cavity equation, the dynamical evolution of the COM system is given by a set of quantum Langevin equations (QLEs):
\begin{equation}\label{eq2}
	\begin{aligned}
\dot{a} &= -(\mathrm{i}\Delta_{\mathrm{fb}}+\kappa_{\mathrm{fb}})a
          + \mathrm{i}g\,a(b^{\dagger}+b)
          + 2G_c a^{\dagger} e^{-\mathrm{i}(\Delta_{c}t-\theta_{c})}
          + E_{\mathrm{fb}}
          + \sqrt{2\kappa_{a}}\,a_{\mathrm{fb}}^{\mathrm{in}},\\
\dot{b} &= -(\mathrm{i}\omega_{b}+\kappa_{b})b
          + \mathrm{i}g\,a^{\dagger}a
          + 2G_m b^{\dagger} e^{-\mathrm{i}(\omega_{m}t-\theta_{m})}
          + \sqrt{2\kappa_{b}}\,b^{\mathrm{in}}.
\end{aligned}
\end{equation}
where
\begin{equation}
\Delta_{\mathrm{fb}}
=
\Delta_{a}
-
\frac{2\kappa_{a}r_{b}\sin\theta}
{1+r_b^2+2r_b\cos\theta},
\qquad
\kappa_{\mathrm{fb}}
=
\kappa_a
\frac{1-r_b^2}
{1+r_b^2+2r_b\cos\theta}
\label{eq4}
\end{equation}
are the effective cavity detuning and decay rate, respectively. The expressions in Eq.~(\ref{eq4}) follow by separating the feedback contribution proportional to $a$ into its real and imaginary parts. The real part modifies the cavity decay rate, while the imaginary part shifts the cavity detuning \cite{Harwod2021,Wei2021}.
The operator $b^{\mathrm{in}}(t)$ and the parameter $\kappa_b$ represent the input noise and the decay rate of the phonon mode, respectively.

The input noise  is characterized by the following correlation functions: 

\begin{equation}
\begin{aligned}
\left\langle
a_{\mathrm{fb}}^{\mathrm{in}}(t)
a_{\mathrm{fb}}^{\mathrm{in}\dagger}(t')
\right\rangle
&=
\kappa_{\mathrm{fb}}/\kappa_{a}
\left[N_a(\omega_a)+1\right]
\delta(t-t'),\\
\left\langle
a_{\mathrm{fb}}^{\mathrm{in}\dagger}(t)
a_{\mathrm{fb}}^{\mathrm{in}}(t')
\right\rangle
&=
\kappa_{\mathrm{fb}}/\kappa_{a}
 N_a(\omega_a)
\delta(t-t'),\\
\left\langle
b^{\mathrm{in}}(t)
b^{\mathrm{in}\dagger}(t')
\right\rangle
&=
\left[N_b(\omega_b)+1\right]
\delta(t-t'),\\
\left\langle
b^{\mathrm{in}\dagger}(t)
b^{\mathrm{in}}(t')
\right\rangle
&=
N_b(\omega_b)
\delta(t-t').
\end{aligned}
\label{eq:noise_correlations}
\end{equation}
Here $N_{o}(\omega_{o})=[\exp(\frac{\hbar\omega_{o}}{k_\textrm{B}T})-1]^{-1} (o=a,b)$ is the mean thermal excitation number at the environment temperature $T$, and $k_\textrm{B}$ is the Boltzmann constant.

We expand each operator as a sum of its steady-state mean value and a small quantum fluctuation, i.e., $o = \langle o \rangle + \delta o$. By substituting this expression into the QLEs and ignoring the second order fluctuation terms, the linearized QLEs of the quantum fluctuations are given as follows:
\begin{eqnarray}
\nonumber\delta\dot{a} &=& -(\mathrm{i}\Delta_{\mathrm{fb}}^{\prime}+\kappa_{\mathrm{fb}})\delta a+\mathrm{i}G_{\mathrm{eff}}(\delta b^{\dagger}+\delta b)+2G_c\delta a^{\dagger}e^{-\mathrm{i}(\Delta_{c}t-\theta_{c})}+\sqrt{2\kappa_a}{a_{\mathrm{fb}}}^{\textrm{in}}(t),\\
\delta\dot{b} &=&-(\mathrm{i}\omega_{b}+\kappa_{b})\delta b+\mathrm{i}(G_{\mathrm{eff}}\delta a^{\dagger}+G_{\mathrm{eff}}^{\ast}\delta a)+2G_{m}\delta b^{\dagger}e^{-\mathrm{i}(\omega_{m}t-\theta_{m})}+\sqrt{2\kappa_{b}}{b}^{\textrm{in}}(t),
\end{eqnarray}
where $\Delta_{\mathrm{fb}}^{\prime}=\Delta_{\mathrm{fb}}-g(\langle b\rangle+\langle b\rangle^*)$, $G_{\mathrm{eff}}(t)=g\langle a(t)\rangle=G_x(t)+\mathrm{i}G_y(t)$ is the effective COM coupling rate. We can rewrite the linearized QLEs in a matrix form by defining the quadrature fluctuation operators, i.e., $X_{o(o^{\mathrm{in}})}(t)=\frac{1}{\sqrt{2}}[o(o^{\mathrm{in}})+o^{\dagger}(o^{\mathrm{in}\dagger})]$ and $Y_{o(o^{\mathrm{in}})}(t)=\frac{1}{\sqrt{2}\mathrm{i}}[o(o^{\mathrm{in}})-o^{\dagger}(o^{\mathrm{in}\dagger})]$.  The matrix form is $\dot{u}(t)=A(t)u(t)+n(t)$, where $u^{T}(t)=[\delta X_={a}(t),\delta Y_{a}(t), \delta X_{b}(t), \delta Y_{b}(t)]$ and~$n^{T}(t)=
[\sqrt{2\kappa_a}X_{a,\mathrm{fb}}^{\mathrm{in}},
\sqrt{2\kappa_a}Y_{a,\mathrm{fb}}^{\mathrm{in}},
\sqrt{2\kappa_b}X_b^{\mathrm{in}},
\sqrt{2\kappa_b}Y_b^{\mathrm{in}}]$. The drift matrix $A(t)$ is given as follows:

\begin{equation}
A(t) = \left( \begin{array}{c c c c }
    -\kappa_{\mathrm{fb}}+\Gamma_{c} & \Delta_{\mathrm{fb}}^{\prime}-\zeta_{c} & -2G_y & 0 \\
   -\Delta_{\mathrm{fb}}^{\prime}-\zeta_{c} & -\kappa_{\mathrm{fb}}-\Gamma_{c} & 2G_x & 0   \\
    0&0 & -\kappa_b+\Gamma_m & \omega_b-\zeta_m \\
    2G_x& 2G_y &-\omega_{b}-\zeta_m &  -\kappa_b-\Gamma_m \\
  \end{array} \right),
\end{equation}
where $\Gamma_{m}=2G_{m}\mathrm{cos}(\omega_{m}t-\theta_{m})$, $\Gamma_{c}=2G_{c}\mathrm{cos}(\Delta_{c}t-\theta_{c})$, $\zeta_{m}=2G_{m}\mathrm{sin}(\omega_{m}t-\theta_{m})$, $\zeta_{c}=2G_{c}\mathrm{sin}(\Delta_{c}t-\theta_{c})$. We find that the system dynamics are affected by OPA, MPA, and the coherent feedback loop. Therefore, we show that the dynamics of the system can be regulated by the reflectivity of the CBS in Fig.~\ref{fig2}.
The dynamics of the quantum fluctuations can be fully characterized by the covariance matrix (CM) $\mathbf{V}$, a $4 \times 4$ real symmetric matrix, with its matrix elements defined as $V_{ij} = \frac{1}{2} \langle u_i(t) u_j(t) + u_j(t) u_i(t) \rangle$,($i, j = 1, 2, 3, 4$). The equation of motion of the CM ($V(t)$) is,
\begin{equation}
\dot{V}(t)=\mathbf{A}(t) \mathbf{V}(t) + \mathbf{V}(t) \mathbf{A}(t)^\mathrm{T}+\mathbf{D}. \label{eq:lyapunov}
\end{equation}

The diffusion matrix $D$ is
\begin{equation}
\begin{aligned}
D=\operatorname{diag}\Bigg[
\kappa_{\mathrm{fb}}(2N_a+1),
\kappa_{\mathrm{fb}}(2N_a+1),\kappa_b(2N_b+1),
\kappa_b(2N_b+1)
\Bigg].
\end{aligned}
\end{equation}

Unless otherwise specified, we set $\theta=0$ in the numerical calculations below.

The CM can be written in block matrix form 
\begin{equation}
V(t) = \left( \begin{array}{c c }
    V_a & V_{ab} \\
    V_{ab}^{T}  &V_b \\
  \end{array} \right),
  \label{eq10}
\end{equation}
where $V_a$, $V_b$, $V_{ab}$ are $2\times2$ subblock matrices corresponding to the photon mode, the phonon mode and their correlation, respectively. 

In order to quantify the bipartite optomechanical entanglement, we use logarithmic negativity $E_{N}$, which is defined as \cite{Adesso2004}
\begin{equation}
E_{N}=\mathrm{max}[0,-\mathrm{ln}{2\eta^{-}})],
\end{equation}
where $\eta^{-}=\frac{1}{\sqrt{2}}\sqrt{\Sigma-\sqrt{\Sigma^2-4\mathrm{det}(V)}}$, $\Sigma=\mathrm{det(V_a)}+\mathrm{det}(V_b)-2\mathrm{det}V_{ab}$.

The steering from the photon mode to the phonon mode, i.e., the photon mode can steer the phonon mode, is defined as
\begin{eqnarray}
\mathcal{G}_{a\rightarrow b}&=&\max[ 0 , \frac{1}{2}\mathrm{ln}\frac{\mathrm{det}V_a}{4\mathrm{det}V}].
\end{eqnarray}
Similarly, the steering from the phonon mode to the photon mode is given by
\begin{eqnarray}
	\mathcal{G}_{b\rightarrow a}&=&\max\left[ 0 , \frac{1}{2}\mathrm{ln}\frac{\mathrm{det}V_b}{4\mathrm{det}V}\right].
\end{eqnarray}

According to Eq.~\eqref{eq2}, the classical mean values $\langle o(t)\rangle$ may evolve toward an asymptotic periodic orbit according to amplitude-modulated driving schemes \cite{Mari2009,Hu2019}, which has been verified in  Fig.~\ref{fig2}. Therefore, the long-time dynamics is periodic in time with some period $\tau$. Then we have $A(t+\tau)=A(t)$ and $V(t+\tau)=V(t)$. Similarly, the entanglement, steering, squeezing degree, and purity will acquire the same periodicity in time as $V(t)$. In this case, we quantify them by their maxima in one period, i.e.,
\begin{eqnarray}
\nonumber E_{N,\mathrm{max}} &=& \underset{\tau}{\max}[E_{N}(t)],\\
\nonumber\mathcal{G}_{a\rightarrow b,\mathrm{max}}&=&\underset{\tau}{\max}[\mathcal{G}_{a\rightarrow b}(t)],\\ 
\mathcal{G}_{b\rightarrow a,\mathrm{max}} &=& \underset{\tau}{\max}[\mathcal{G}_{b\rightarrow a}(t)].
 \end{eqnarray}

\section{\label{sec:level3}Numerical simulations and discussion}
In this section, we numerically investigate how quantum correlations can be controlled through the combined effects of OPA, MPA, and coherent feedback. We study how coherent feedback and the OPA–MPA modulation-frequency and pump-amplitude ratios affect optomechanical entanglement. Moreover, we analyze the mechanisms responsible for the enhancement of these quantum correlations and quantitatively compare the combined scheme with several reference configurations, particularly the system with dual parametric amplification but without coherent feedback. Finally, we examine the control of entanglement and EPR steering through the feedback phase and assess their robustness against thermal noise. We choose experimentally feasible parameters within the range of typical values \cite{yang2024,Shao2024}:  $\omega_b/2\pi=1~\mathrm{MHz}$,  $\kappa_a/2\pi= 0.5~\mathrm{MHz}$, $\kappa_b/2\pi=1~\mathrm{Hz}$, $E/2\pi=65~\mathrm{GHz}$, $\lambda_{l}=1550~\mathrm{nm}$, $P_{l}\approx 2.9~\mathrm{mW},$ $m=150~\mathrm{ng}$, $g/2\pi=4~\mathrm{Hz}$ and $T=20~\mathrm{mK}$. Throughout the numerical calculations, dynamical stability is assessed from the Floquet multipliers, and only stable parameter points are retained (see Appendix~\ref{AppendixA}).

\begin{figure}[htbp]
  \begin{center}
    \includegraphics[width=\linewidth]{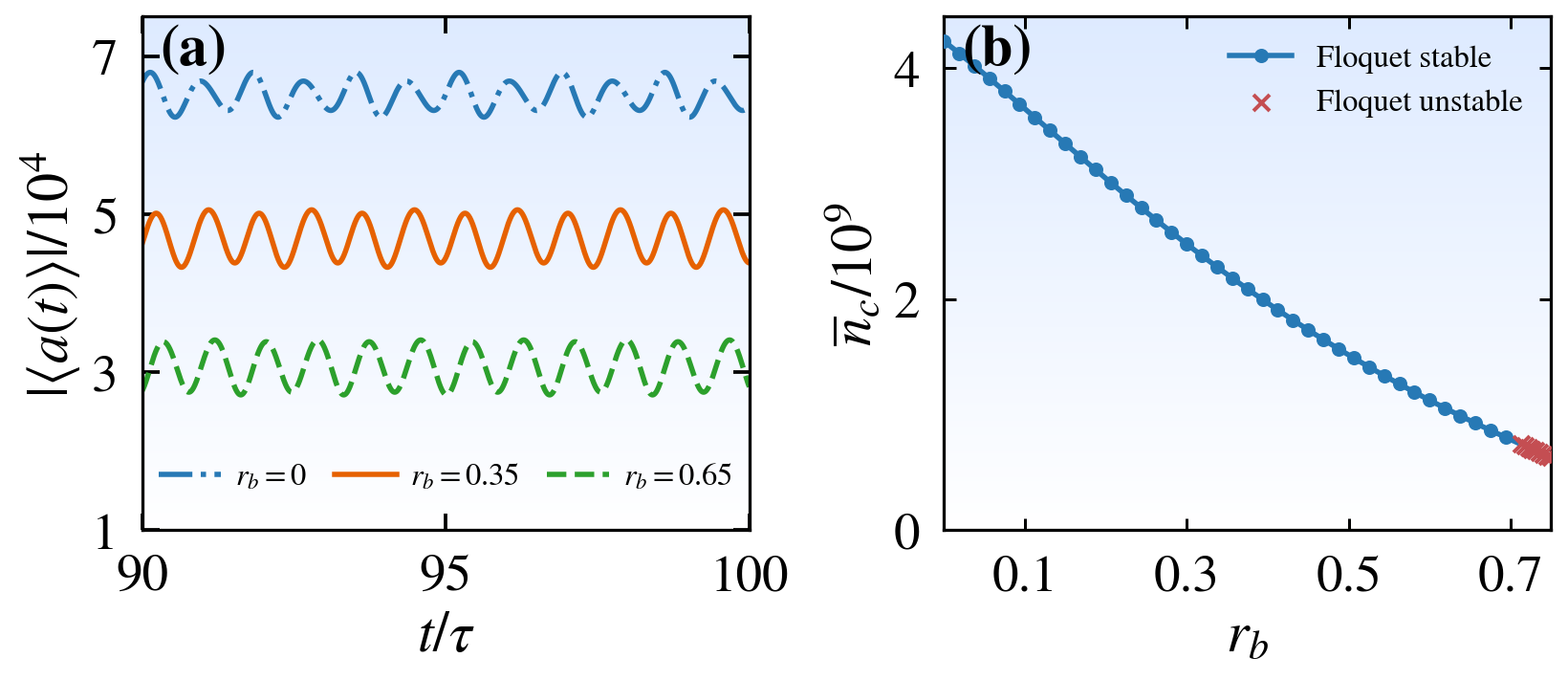} 
  \caption{Feedback dependence of the intracavity field dynamics and photon number. (a) Long-time evolution of the mean cavity-field amplitude $|\langle a(t)\rangle|$ for $r_b=0$, $0.35$, and $0.65$. (b) Period-averaged intracavity photon number as a function of $r_b$. Here, $G_m=0.04\omega_b, G_c=0.01\omega_b$, $\Delta_a=\omega_b$, $\Delta_c=1.18\omega_b$, $\omega_m=1.5\Delta_c$.}
\label{fig2}
   \end{center}
  \end{figure}

Figure~\ref{fig2}(a) shows that the cavity-field amplitude exhibits periodic oscillations after the transient evolution. As $r_b$ increases, the oscillations shift toward smaller values of $|\langle a(t)\rangle|$, consistent with the monotonic decrease of the period-averaged intracavity photon number shown in Fig.~\ref{fig2}(b).
For $\theta=0$, the analysis in Appendix~\ref{AppendixB} gives $d n_c^{\mathrm{lin}}/d r_b<0$ when $|\Delta_a|>\kappa_{\mathrm{fb}}$. This condition is satisfied throughout the parameter range considered here. Consequently, the intracavity photon number decreases with $r_b$ despite the simultaneous reduction of $\kappa_{\mathrm{fb}}$. As shown below, entanglement and steering are nevertheless enhanced as the intracavity photon number calculated from the mean field and averaged over one period decreases, suggesting that the feedback-induced enhancement does not originate from a larger intracavity photon number.

\begin{figure}[htbp]
	\begin{center}
	\includegraphics[width=\linewidth]{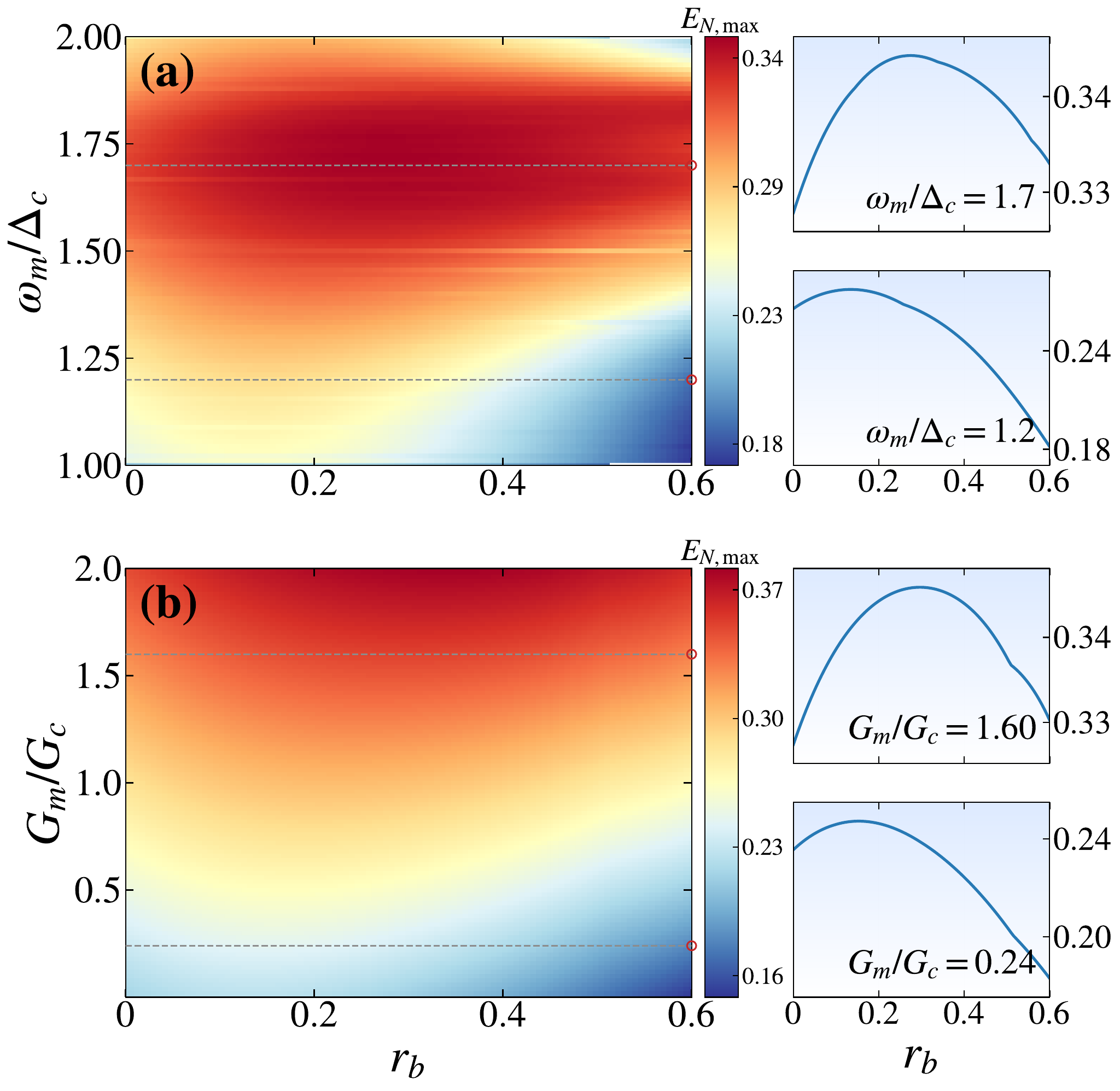}
		\caption{Maximum entanglement $E_{N,\mathrm{max}}$ as functions of $\omega_m/\Delta_c$ and $r_b$ in (a), and as functions of $G_m/G_c$ and $r_b$ in (b). Here, $G_c=0.01\,\omega_b$ in both panels, with $G_m/G_c=1.5$ in (a) and $\omega_m/\Delta_c=1.6$ in (b). The other parameters are the same as those in Fig.~\ref{fig2}.}
		\label{fig3}
	\end{center}
\end{figure}

Figure~\ref{fig3}(a) shows $E_{N,\max}$ as a function of $\omega_m/\Delta_c$ and $r_b$. Relatively large values of $E_{N,\max}$ are obtained over a broad parameter region around $\omega_m/\Delta_c\simeq1.7$. This enhancement originates from the mechanical parametric resonance $\omega_m\simeq2\omega_b$. In the interaction picture, the MPA term contains the factor $\exp[-\mathrm{i}(\omega_m-2\omega_b)t]$ and therefore resonantly enhances mechanical squeezing near this region~\cite{Bothner2020}. The slight shift of the optimum away from exact resonance arises from the combined effects of the OPA, MPA, optomechanical interaction, coherent feedback, and dissipation. The line cuts in the right-hand panels, corresponding to the dashed lines in Fig.~\ref{fig3}(a), reveal a nonmonotonic dependence of $E_{N,\max}$ on $r_b$.
 Increasing $r_b$ simultaneously reduces $\kappa_{\mathrm{fb}}$ and the effective driving amplitude. The reduced cavity decay rate lengthens the cavity lifetime and initially enhances $E_{N,\max}$, whereas the weakened drive reduces $\alpha(t)$ and hence the linearized optomechanical coupling $G(t)=g\alpha(t)$. At sufficiently large $r_b$, the latter effect becomes dominant. Because the MPA effect is stronger near the mechanical parametric resonance, the maximum for $\omega_m/\Delta_c=1.7$ occurs at a larger $r_b$ and is followed by a more gradual decrease than that for $\omega_m/\Delta_c=1.2$. Figure~\ref{fig3}(b) shows $E_{N,\max}$ as a function of $G_m/G_c$ and $r_b$ at $\omega_m/\Delta_c=1.6$. For a fixed $r_b$, $E_{N,\max}$ generally increases with $G_m/G_c$ owing to the stronger mechanical parametric amplification. Increasing $G_m/G_c$ also shifts the optimal $r_b$ to a larger value and broadens the parameter range over which large $E_{N,\max}$ is obtained. At sufficiently large $r_b$, however, the reduction in the effective drive and linearized optomechanical coupling again becomes dominant.

\begin{figure}[htbp]
	\begin{center}
	\includegraphics[width=1\linewidth]{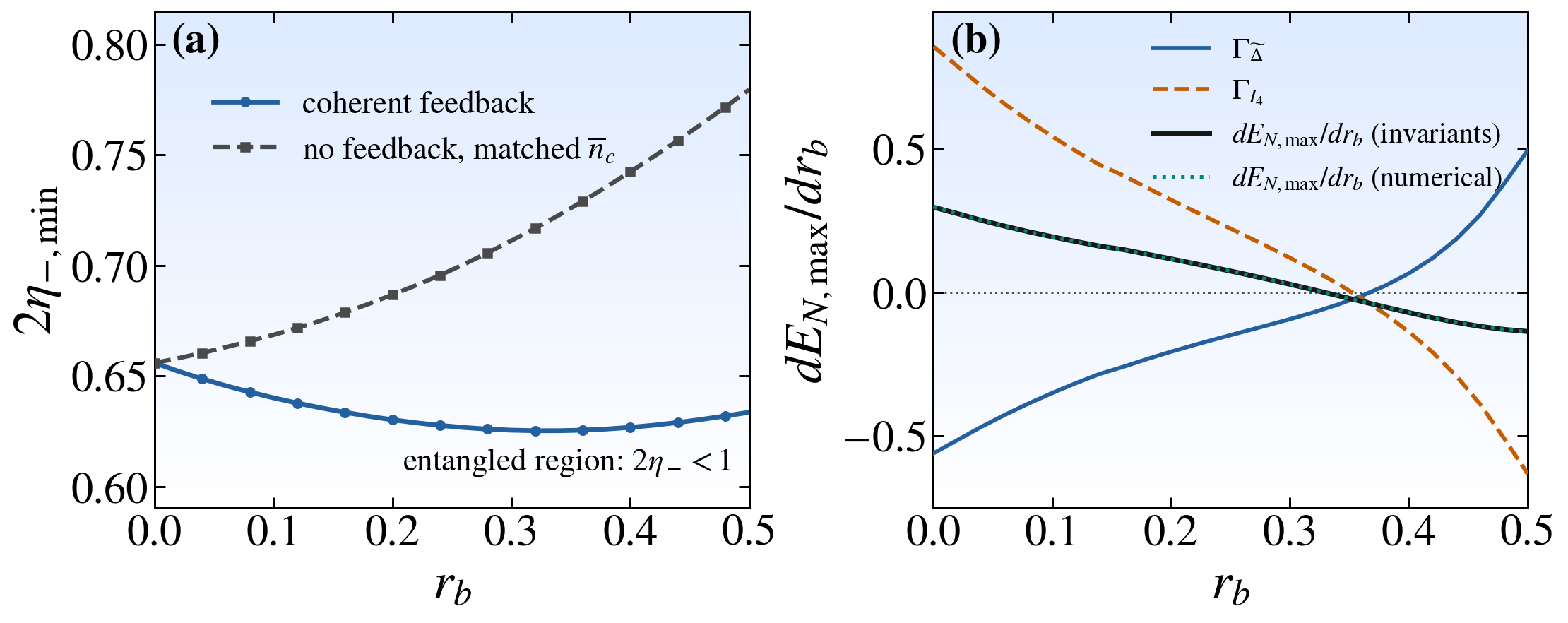}
		\caption{Feedback enhancement of optomechanical entanglement at the same period-averaged intracavity photon number. (a) $2\eta_{-,\min}$ for the coherent-feedback system and for the no-feedback reference with matched $\bar n_c$. (b) Direct numerical derivative $dE_{N,\max}/dr_b$ and its invariant contributions $\Gamma_{\tilde{\Delta}}$ and $\Gamma_{I_4}$. The parameters are the same as those in Fig.~\ref{fig2}.}
		\label{fig4}
	\end{center}
\end{figure}

To determine whether the enhancement observed in Fig.~\ref{fig3} results simply from a change in the intracavity photon number, we compare the feedback system with a system without feedback at the same value of $\overline{n}_{c}$. For each value of $r_b$, the driving amplitude of the system without feedback is adjusted until its $\overline{n}_{c}$ equals that of the feedback system. As shown in Fig.~\ref{fig4}(a), the two systems exhibit opposite trends. The value of $2\eta_{-,\min}$ decreases in the presence of feedback but increases in the matched system without feedback. Because a smaller $2\eta_{-,\min}$ corresponds to a larger logarithmic negativity, this comparison shows that the feedback-induced enhancement persists at matched $\bar n_c$ and therefore cannot be attributed to an increase in the period-averaged intracavity photon number.
To characterize the covariance-matrix changes underlying this behavior, Fig.~\ref{fig4}(b) decomposes $dE_{N,\max}/dr_b$ into the contributions $\Gamma_{\widetilde{\Delta}}$ and $\Gamma_{I_4}$ associated with $\widetilde{\Delta}$ and $I_4=\det V$, respectively. At small $r_b$, the positive contribution $\Gamma_{I_4}$ outweighs the negative contribution $\Gamma_{\widetilde{\Delta}}$, yielding $dE_{N,\max}/dr_b>0$. Near $r_b\simeq0.35$, the two contributions cancel and $E_{N,\max}$ reaches its maximum. At larger $r_b$, $\Gamma_{I_4}<0$ outweighs $\Gamma_{\widetilde{\Delta}}>0$, so that $dE_{N,\max}/dr_b<0$ and the entanglement decreases (See Appendix~\ref{AppendixC}). The agreement between $\Gamma_{\widetilde{\Delta}}+\Gamma_{I_4}$ and the direct numerical derivative validates the invariant decomposition. Together with the photon-number-matched comparison in Fig.~\ref{fig4}(a), this result shows that coherent feedback enhances entanglement by modifying the fluctuation and correlation structure encoded in $V$, rather than by increasing $\bar n_c$.

\begin{figure}[htbp]
	\begin{center}
	\includegraphics[width=1\linewidth]{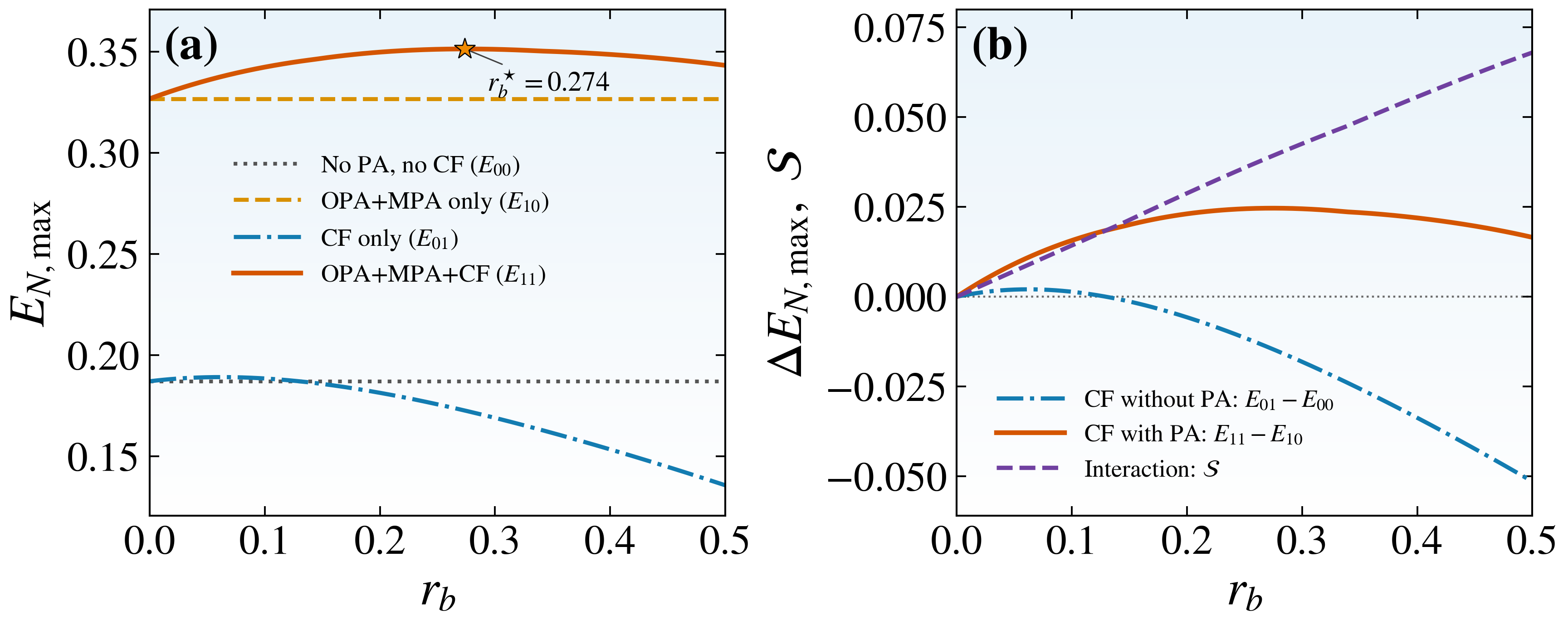}
		\caption{Quantitative comparison of the four control configurations at $\omega_m/\Delta_c=1.7$. For the configurations with dual parametric amplification ($i=1$), $G_m/G_c=1.5$. (a) $E_{N,\max}$ versus $r_b$ for
the reference system ($E_{00}$), dual parametric amplification alone ($E_{10}$), coherent feedback alone ($E_{01}$), and their combined
configuration ($E_{11}$). The star marks the maximum of $E_{11}$ over the
displayed range. (b) Feedback-induced changes without and with dual
parametric amplification, $E_{01}-E_{00}$ and $E_{11}-E_{10}$,
respectively, together with the nonadditive interaction contrast
$\mathcal{S}=E_{11}-E_{10}-E_{01}+E_{00}$.} 
		\label{fig5}
	\end{center}
\end{figure}

\begin{table}[t]
\caption{Two-factor comparison at $r_b^\star=0.274$, where
$E_{11}$ reaches its maximum on the displayed grid. PA denotes the
simultaneous use of the OPA and MPA. Here,
$\Delta_{\mathrm{CF}1}=E_{01}-E_{00}$ without PA and
$\Delta_{\mathrm{CF}2}=E_{11}-E_{10}$ with PA.}
\label{tab1}
\begin{ruledtabular}
\begin{tabular}{lcccc}
Configuration & PA & CF & $E_{N,\max}$ & $\Delta_{\mathrm{CF}(i)}$ \\
\hline
Reference system, $E_{00}$ & off & off & 0.1871 & -- \\
PA only, $E_{10}$ & on & off & 0.3268 & -- \\
CF only, $E_{01}$ & off & on & 0.1726 & $-0.0145$ \\
PA+CF, $E_{11}$ & on & on & 0.3514 & $+0.0246$ \\
\end{tabular}
\end{ruledtabular}
\end{table}

To quantify the interplay between dual parametric amplification and coherent feedback, we perform a two-factor comparison, as shown in Fig.~\ref{fig5}. We denote the value of $E_{N,\max}$ for each configuration by $E_{ij}$, where $i=0$ ($1$) indicates that both the OPA and MPA are switched off (on), while $j=0$ ($1$) denotes the absence (presence) of coherent feedback. All four configurations are evaluated using the same input driving amplitude, bare detunings, damping rates, and thermal parameters; only dual parametric amplification and coherent feedback are switched on or off according to the configuration.
 For the configurations without coherent feedback ($j=0$), $r_b$ does not enter the system, and $E_{00}$ and $E_{10}$ are therefore shown as horizontal reference lines.

Table~\ref{tab1} summarizes the two-factor
comparison of the four configurations. At $r_b^\star=0.274$, the four configurations yield
$E_{00}=0.1871$, $E_{10}=0.3268$, $E_{01}=0.1726$, and
$E_{11}=0.3514$. Interestingly, coherent feedback alone reduces $E_{N,\max}$ by
$E_{00}-E_{01}=0.0145$, whereas applying the same
feedback to the dual-parametric-amplification system increases it by
$E_{11}-E_{10}=0.0246$. To quantify this configuration-dependent
feedback response, we define the nonadditive interaction contrast
\begin{equation}
\mathcal{S}(r_b)
=
E_{11}(r_b)-E_{10}-E_{01}(r_b)+E_{00}.
\label{eq15}
\end{equation}
At $r_b^\star$, we obtain $\mathcal{S}=0.0391>0$. According to Eq.~(\ref{eq15}), $\mathcal{S}$ compares the changes in $E_{N,\max}$ caused by coherent feedback with and without dual parametric amplification. Its positive value therefore indicates that coherent feedback produces a larger change in $E_{N,\max}$ when dual parametric amplification is present. This dependence of the feedback response on the configuration reveals a positive nonadditive interplay, showing that the combined result cannot be interpreted as a simple sum of two independent contributions.

\begin{figure}[htbp]
	\begin{center}
	\includegraphics[width=0.6\linewidth]{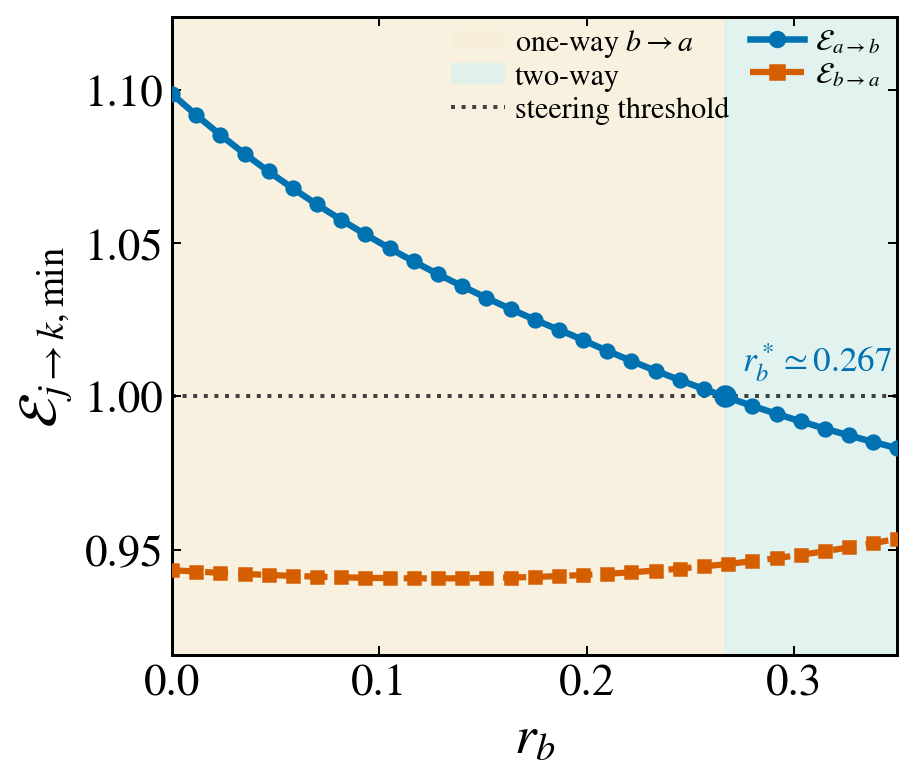}
		\caption{Feedback-controlled transition from one-way to two-way EPR steering. The other parameters are the same as those in Fig.~\ref{fig4}.}
		\label{fig6}
	\end{center}
\end{figure}

Having established that the entanglement enhancement cannot be explained simply by an increase in the period-averaged intracavity photon number, we now examine the directionality of EPR steering. As shown in Fig.~\ref{fig6}, in the absence of feedback, the minimum conditional noises over one modulation period satisfy $\mathcal{E}_{b\to a,\min}<1$, whereas $\mathcal{E}_{a\to b,\min}>1$. Hence, the system initially exhibits one-way phonon-to-photon steering. As $r_b$ increases, $\mathcal{E}_{a\to b,\min}$ decreases and crosses the steering threshold at $r_b^*\simeq0.267$, while $\mathcal{E}_{b\to a,\min}$ remains below unity. Coherent feedback therefore enables photon-to-phonon steering while preserving phonon-to-photon steering, leading to two-way EPR steering for $r_b>r_b^*$.
 As detailed in Appendix~\ref{AppendixD}, the two directional conditional noises are determined by different Schur complements of the covariance matrix, which account for their different responses to coherent feedback.

\begin{figure}[htbp]
  \begin{center}
    \includegraphics[width=\linewidth]{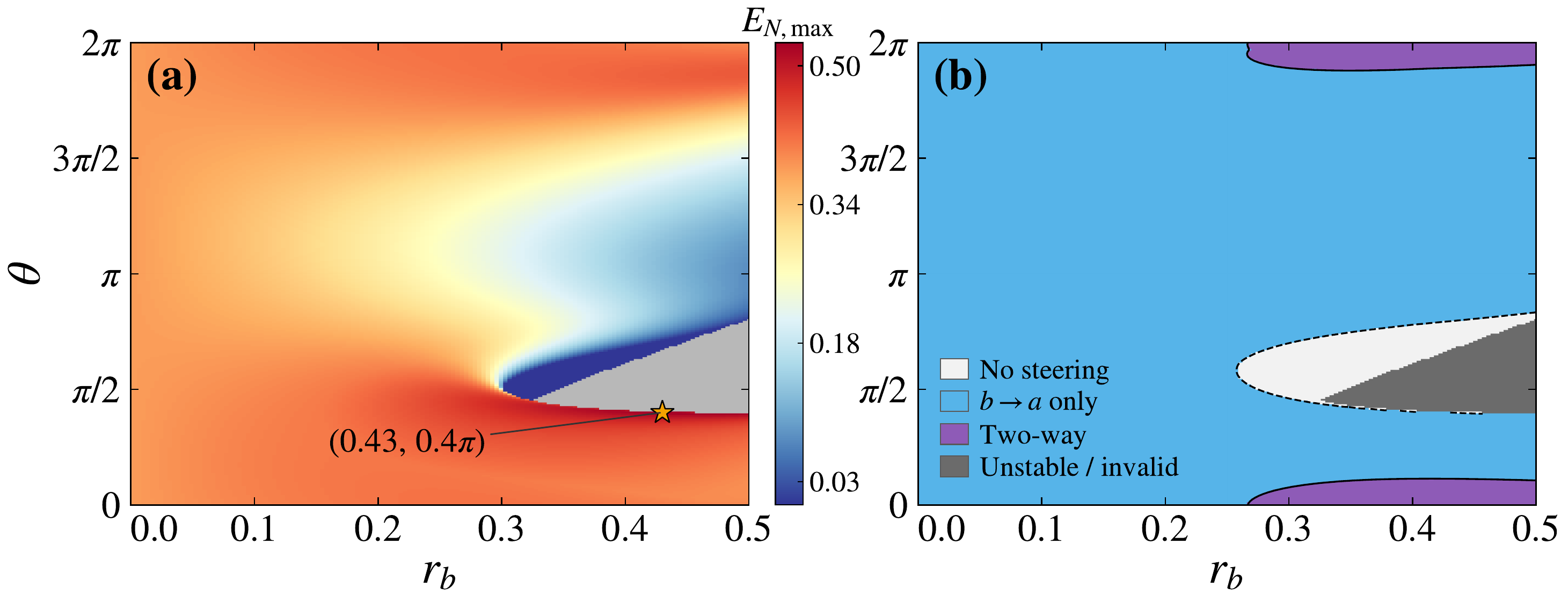} 

\caption{Joint dependence of optomechanical entanglement and Gaussian EPR steering on the CBS reflectivity and the feedback phase. (a) $E_{N,\max}$ as a function of $r_b$ and $\theta$. The star marks $(r_b,\theta)=(0.43,0.4\pi)$. (b) Steering regimes in the $(r_b,\theta)$ plane, with the colors defined in the panel. The colors identify no steering, one-way phonon-to-photon steering, and two-way steering, as indicated in the legend. Gray denotes excluded parameter points at which the periodic solution is Floquet unstable or the numerically obtained covariance matrix fails the physicality condition. The remaining parameters in panels (a) and (b) are the same as those in Figs.~\ref{fig3}(a) and \ref{fig6} respectively.}
\label{fig7}
   \end{center}
  \end{figure}

To further clarify the role of the coherent feedback phase, Fig.~\ref{fig7} shows the joint dependence of the quantum correlations on the CBS reflectivity $r_b$ and the feedback phase $\theta$. Fig.~\ref{fig7}(a) demonstrates that $E_{N,\max}$ depends strongly and nonmonotonically on both feedback parameters. This behavior arises because changing $\theta$ modifies the interference between the incident field and the field returned through the feedback loop. Consequently, the effective cavity drive, decay rate, and detuning, and hence the periodic intracavity field and linearized optomechanical interaction, are modified simultaneously. Increasing $r_b$ therefore does not enhance the entanglement for every feedback phase. Instead, substantial entanglement occurs within phase-dependent stable regions and may be suppressed when the same feedback amplitude reflectivity is combined with a different phase. The star marks the feedback setting $(r_b,\theta)=(0.43,0.4\pi)$. Its location within a finite region of enhanced entanglement shows that this enhancement is not restricted to an isolated point in the $(r_b,\theta)$ plane.
The gray regions correspond to parameter values for which the linearized periodic dynamics are Floquet unstable or the resulting covariance matrix violates the Robertson–Schrödinger uncertainty relation. They are therefore excluded from the analysis. The phase-dependent boundaries of these regions also show that the feedback phase provides an important constraint on the accessible stable operating range. Figure~\ref{fig7}(b) maps the steering regimes in the $(r_b,\theta)$ plane, with all other parameters as in Fig.~\ref{fig6}. At $\theta=0$ (equivalently, $\theta=2\pi$), the result reproduces that of Fig.~\ref{fig6}: $r_b^*\simeq0.267$ marks the transition from one-way phonon-to-photon steering to two-way steering. Away from $\theta=0$, the boundaries between the different steering regimes shift with the feedback phase. Thus, the joint tuning of $r_b$ and $\theta$ enables switching among no steering, one-way phonon-to-photon steering, and two-way steering over finite parameter regions.

\begin{figure}[htbp]
  \begin{center}
    \includegraphics[width=\linewidth]{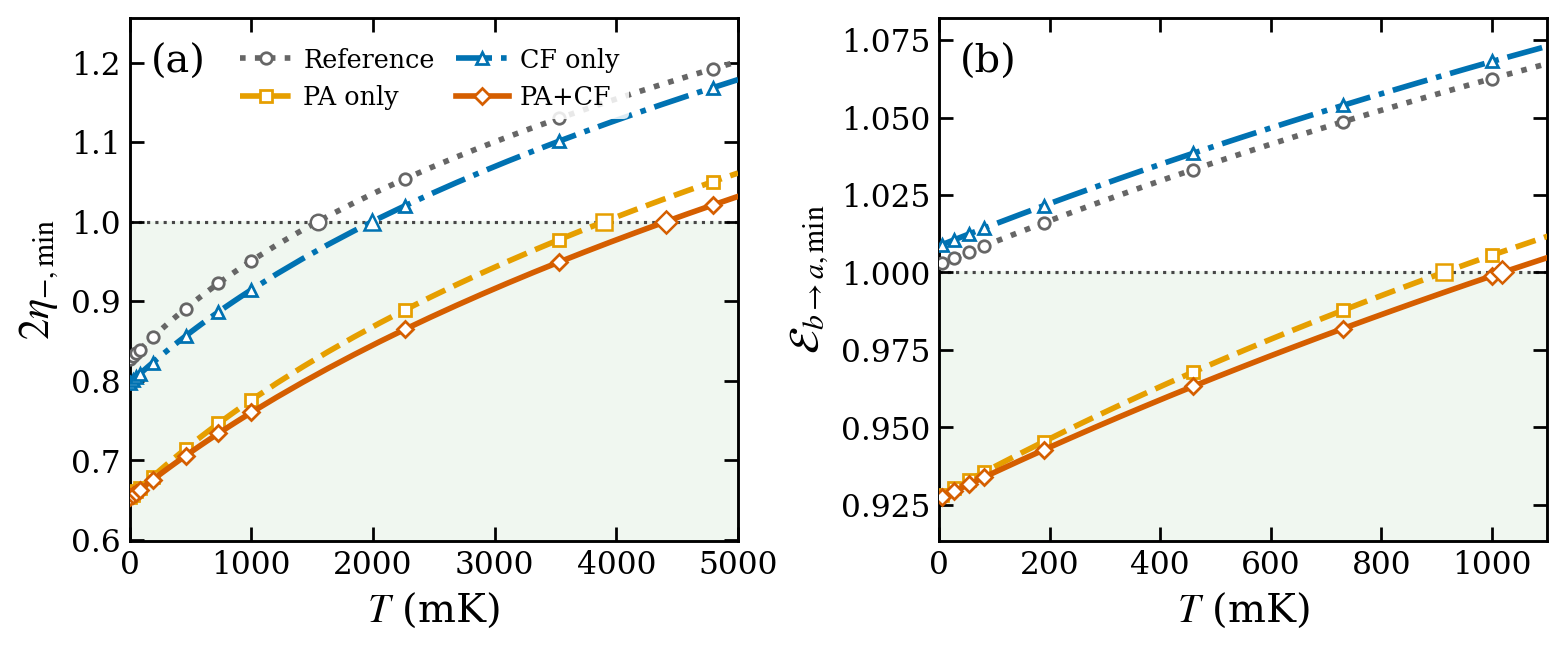} 

\caption{Thermal robustness of optomechanical entanglement and phonon-to-photon EPR steering for the four configurations. Here, $r_b=0.11$ and $\theta=0.575\pi$. The other parameters are the same as those in Fig.~\ref{fig7}.}
\label{fig8}
   \end{center}
  \end{figure}
  
Finally, we examine the robustness of optomechanical entanglement and phonon-to-photon EPR steering against thermal noise. Figure \ref{fig8} shows $2\eta_{-,\min}$ and $\mathcal E_{b\rightarrow a,\min}$ as functions of the mechanical bath temperature $T$ for the reference configuration, dual parametric amplification alone, coherent feedback alone, and the combined configuration. As $T$ increases, the mean thermal occupation of the mechanical bath and the corresponding mechanical diffusion term increase. Consequently, both plotted quantities increase and the associated quantum correlations weaken. For a curve that is initially below unity, its crossing of unity defines the threshold temperature above which the corresponding criterion is no longer satisfied. Figure~\ref{fig8}(a) shows that entanglement persists up to $1.54~\mathrm{K}$ for the reference configuration and $1.99~\mathrm{K}$ for coherent feedback alone. Dual parametric amplification increases this threshold to $3.9~\mathrm{K}$, while the combined configuration further increases it to $4.4~\mathrm{K}$. Figure~\ref{fig8}(b) shows that the phonon-to-photon steering criterion remains above unity for the reference configuration and coherent feedback alone throughout the temperature range considered. With dual parametric amplification, phonon-to-photon steering persists up to $0.91~\mathrm{K}$, and the combined configuration increases this threshold to $1.02~\mathrm{K}$. Thus, at this working point, phonon-to-photon steering is more sensitive to temperature than entanglement, while adding coherent feedback to dual parametric amplification increases the threshold temperature for both correlations.

\FloatBarrier

\section{\label{sec:level4}Experimental Feasibility And Applications}

In this section, we discuss the experimental feasibility of the proposed COM system. It consists of the nonlinear medium and the membrane embedded in the FP cavity assisted by the coherent feedback loop. The OPA and MPA employed in our system provide effective means to enhance the nonlinearity and enrich the quantum dynamics, and can be implemented as follows.

The OPA has been well developed in both theoretical studies and experimental implementations \cite{Hu2019,agarwal2016strong,huang2017robust,he2022force,singh2023enhanced,marhic2015fiber,pan2018experimental,zhang2023optical}. It can be realized by placing a PPKTP crystal inside the cavity and driving it with a classical pump field \cite{vahlbruch2008observation,vahlbruch2016detection,schnabel2017squeezed}. Its effective strength can be tuned by the pump power, whereas the corresponding squeezing phase can be adjusted through the relative phase locking among the pump, signal, and local-oscillator fields \cite{vahlbruch2008observation,vahlbruch2016detection}.

The MPA can then be activated by a periodic piezoelectric or capacitive drive that modulates the effective spring constant of a Si$_3$N$_4$ membrane. Taking the experimental control of the MPA amplitude and phase as an example, Mashaal \textit{et al.} recently investigated parametric modulation in high-stress Si$_3$N$_4$ membrane resonators using both piezoelectric and capacitive schemes. In the piezoelectric case, the MPA amplitude was tuned by increasing the parametric drive voltage $V_p$, with the modulation strength characterized by $g_s=V_p/V_t$. In the capacitive case, the applied voltage $V(t)$ modulated the effective elastic constant $k_p$, which in turn generated the parametric amplification. Moreover, the MPA phase was set by the relative phase $\phi_p$ between the parametric modulation and the phase-locked reference signal \cite{mashaal2025strong,Bothner2020}.

To detect and verify the entanglement and steering, it is necessary to reconstruct the elements of the reduced covariance matrix \cite{li2018magnon,palomaki2013entangling}. The cavity field quadratures can be directly accessed by homodyne or heterodyne detection of the cavity output \cite{barzanjeh2019stationary,chen2020entanglement}. The mechanical quadratures cannot be directly measured but can be inferred with the help of a sufficiently weak auxiliary probe field whose backaction is negligible. When the probe field, with frequency $\omega_{p,l}$, is red detuned from the resonance frequency $\omega_{p,c}$ of the auxiliary probe cavity by the mechanical frequency $\omega_b$, i.e., $\omega_{p,l}\approx\omega_{p,c}-\omega_b$, the linearized interaction between the probe cavity mode and the mechanical mode reduces to the beam-splitter form, which maps the mechanical state onto the probe output field. In this way, the CM elements associated with the mechanical mode can be reconstructed from the correlations of the probe output field \cite{palomaki2013entangling,jiao2020nonreciprocal,ockeloen2018stabilized}.

The finite propagation time in the feedback loop must also be considered. As shown in Appendix~\ref{Appendix E}, the delay can be neglected when the three dimensionless corrections defined in Eq.~\eqref{eq:delay-small-parameters} are much smaller than unity. For a total feedback-loop optical path length of $L=10~\mathrm{cm}$ and the full stable parameter range used in our calculations, their maximum value is less than $1.7\times10^{-2}$. The finite-delay corrections are therefore negligible, and the instantaneous-feedback approximation is valid for the configuration considered here.

\section{\label{sec:level5}Conclusion}
In conclusion, we have investigated the coherent control of optomechanical entanglement and directional EPR steering in a periodically driven COM system combining OPA and MPA with a coherent feedback loop. Using Floquet stability analysis and the periodic steady-state covariance matrix, we have characterized the stable periodic dynamics and the associated quantum correlations. A comparison performed at the same period-averaged intracavity photon number demonstrates that the feedback-induced enhancement of entanglement persists and therefore cannot be attributed simply to an increase in this quantity. Moreover, comparing the reference system with the individual and combined control schemes reveals a nonadditive interplay between dual parametric amplification and coherent feedback: at the selected operating point, coherent feedback alone suppresses the entanglement, whereas it enhances the entanglement when both parametric amplifications are present.

Within the identified stable regime, tuning the feedback strength and phase enables effective control of optomechanical entanglement and directional EPR steering, including a transition from one-way to two-way steering. The temperature-dependent analysis further shows that the combined scheme improves the robustness of both correlations against thermal noise. These findings may provide a useful route toward robust quantum-state engineering in COM platforms, with potential applications in continuous-variable quantum information processing.

\section{\label{sec:level6}Acknowledgments}

Mei Zhang thanks Junzhong Yang for helpful discussions. Jinhao Jia thanks Lijun He for helpful discussions. Yingru Li thanks Xinchun Li and Juan Yao for helpful discussions. This work was supported by the National Natural Science Foundation of China under Grant No. 11475021 and the National Key Basic Research Program of China under Grant No. 2013CB922000.

\clearpage

\appendix

\section{\label{AppendixA}Floquet stability and the periodic covariance matrix}

Because the OPA and the MPA give rise to explicitly time-periodic coefficients in the drift matrix, the stability of the system cannot, in general, be determined from the instantaneous eigenvalues of $A(t)$. We therefore employ Floquet theory together with the periodic Lyapunov method~\cite{Mari2009,Iivari2020}.

For the positive and commensurate modulation frequencies considered in this work, the frequency ratio can be written in lowest terms as
\begin{equation}
\frac{\omega_m}{\Delta_c}
=
\frac{p}{q},
\label{eq:commensurate-ratio}
\end{equation}
where $p$ and $q$ are coprime positive integers. The fundamental common period is then
\begin{equation}
\tau
=
\frac{2\pi q}{|\Delta_c|}
=
\frac{2\pi p}{|\omega_m|}.
\label{eq:common-period}
\end{equation}
All modulation-frequency ratios sampled in this work are commensurate. For the retained parameter points, the mean fields converge to a $\tau$-periodic asymptotic orbit, and the drift matrix evaluated along this orbit satisfies
\begin{equation}
A(t+\tau)=A(t).
\label{eq:periodic-drift}
\end{equation}

The fundamental matrix associated with the homogeneous fluctuation dynamics is determined by
\begin{equation}
\dot{\Phi}(t)
=
A(t)\Phi(t),
\qquad
\Phi(0)=I_4.
\label{eq:fundamental-matrix}
\end{equation}
The monodromy matrix and its spectral radius are
\begin{equation}
M=\Phi(\tau),
\qquad
\rho(M)=\max_j|\mu_j|,
\label{eq:monodromy}
\end{equation}
where $\mu_j$ are the Floquet multipliers. The periodic linearized system is asymptotically stable if and only if
\begin{equation}
\rho(M)<1.
\label{eq:floquet-stability}
\end{equation}

To construct the periodic covariance matrix, we simultaneously solve
\begin{equation}
\dot{W}(t)
=
A(t)W(t)
+
W(t)A^{T}(t)
+
D,
\qquad
W(0)=0,
\label{eq:diffusion-propagator}
\end{equation}
where $D$ is the diffusion matrix defined in the main text. The diffusion accumulated over one common period is
\begin{equation}
Q_{\tau}=W(\tau).
\label{eq:period-diffusion}
\end{equation}
When Eq.~\eqref{eq:floquet-stability} is satisfied, the covariance matrix at the selected reference phase is the unique solution of the discrete Lyapunov equation
\begin{equation}
V_0
=
MV_0M^{T}
+
Q_{\tau}.
\label{eq:discrete-lyapunov}
\end{equation}
The asymptotic covariance matrix within one common period is then
\begin{equation}
V_{\mathrm p}(t)
=
\Phi(t)V_0\Phi^{T}(t)
+
W(t),
\label{eq:periodic-covariance}
\end{equation}
which satisfies
\begin{equation}
V_{\mathrm p}(t+\tau)
=
V_{\mathrm p}(t).
\label{eq:covariance-periodicity}
\end{equation}

All quantum-correlation measures reported in the main text are evaluated from $V_{\mathrm p}(t)$ using their previously defined expressions. The reported maximum or minimum values are obtained by taking the corresponding extrema over one asymptotic common period, $0\leq t<\tau$.

Only parameter points for which the mean fields converge to a periodic steady state and which satisfy the Floquet stability condition in Eq.~\eqref{eq:floquet-stability} are retained. All covariance matrices used to evaluate the quantum-correlation measures are also verified to satisfy the Robertson--Schrödinger uncertainty relation.

\section{\label{AppendixB}Feedback dependence of the intracavity photon number}

In this appendix, we explain why, for the parameters used in Fig.~2 at $\theta=0$, the period-averaged mean-field intracavity photon number decreases as the feedback amplitude reflectivity $r_b$ increases. Taking the expectation values of the QLEs, factorizing operator products at the mean-field level, and using the zero mean of the input noises, we obtain
\begin{align}
\dot{\alpha}
={}&-(\kappa_{\mathrm{fb}}+\mathrm{i}\Delta_{\mathrm{fb}})\alpha
+\mathrm{i}g\alpha(\beta+\beta^{*})
+2G_c\alpha^{*}e^{-\mathrm{i}(\Delta_c t-\theta_c)}
+E_{\mathrm{fb}},
\nonumber \\
\dot{\beta}
={}&-(\kappa_b+\mathrm{i}\omega_b)\beta
+\mathrm{i}g|\alpha|^2
+2G_m\beta^{*}e^{-\mathrm{i}(\omega_m t-\theta_m)},
\label{eq:A2}
\end{align}
where $\alpha=\langle a\rangle$ and $\beta=\langle b\rangle$. The
feedback-modified parameters are
\begin{align}
\kappa_{\mathrm{fb}}
&=
\kappa_a
\frac{1-r_b^2}
{1+r_b^2+2r_b\cos\theta},
\nonumber\\
\Delta_{\mathrm{fb}}
&=
\Delta_a-
\frac{2\kappa_a r_b\sin\theta}
{1+r_b^2+2r_b\cos\theta},
\nonumber\\
E_{\mathrm{fb}}
&=
\frac{t_bE}{1+r_be^{\mathrm{i}\theta}},
\qquad
t_b=\sqrt{1-r_b^2}.
\label{eq:A3}
\end{align}

The dependence of $E_{\mathrm{fb}}$ on $r_b$ follows from the coherent
superposition of the incident field and the field returned by the feedback
loop. In particular,
\begin{equation}
|E_{\mathrm{fb}}|^2
=E^2\frac{1-r_b^2}
{1+r_b^2+2r_b\cos\theta}.
\label{eq:A4}
\end{equation}
The term $2r_b\cos\theta$ in the denominator is the interference term.
With the input-output convention adopted in this work, $\theta=0$
corresponds to partial destructive interference at the cavity input. Hence,
\begin{equation}
|E_{\mathrm{fb}}|^2
=E^2\frac{1-r_b}{1+r_b},
\label{eq:A5}
\end{equation}
which decreases as $r_b$ increases. Thus, increasing $r_b$ does not necessarily increase the effective cavity drive. Although it increases the fraction of the output field recirculated through the feedback loop, it simultaneously reduces the beam-splitter amplitude transmission $t_b$ and strengthens the destructive interference with the incident field at $\theta=0$.

To determine whether the reduction in $|E_{\mathrm{fb}}|^2$ is compensated by the simultaneous decrease in the cavity decay rate, we use the feedback-dependent linear cavity response as a reference. The OPA, MPA, and optomechanical terms are omitted only from this analytical estimate but are retained in the full numerical calculation of the periodic dynamics shown in Fig.~\ref{fig2}(a). The stationary response of this linear reference model satisfies\begin{equation}
0=-(\kappa_{\mathrm{fb}}+\mathrm{i}\Delta_{\mathrm{fb}})\alpha
+E_{\mathrm{fb}},
\label{eq:A6}
\end{equation}
and therefore, within the same approximation,
\begin{equation}
n_c^{\mathrm{lin}}
\equiv
|\alpha|^2
=
\frac{|E_{\mathrm{fb}}|^2}
{\kappa_{\mathrm{fb}}^2+\Delta_{\mathrm{fb}}^2}.
\label{eq:A6}
\end{equation}

For $\theta=0$, the feedback parameters reduce to
\begin{equation}
\kappa_{\mathrm{fb}}
=\kappa_a\frac{1-r_b}{1+r_b},
\qquad
\Delta_{\mathrm{fb}}=\Delta_a,
\qquad
E_{\mathrm{fb}}
=E\sqrt{\frac{1-r_b}{1+r_b}}.
\label{eq:A8}
\end{equation}
Substituting Eq.~(\ref{eq:A8}) into Eq.~(\ref{eq:A6}) gives
\begin{equation}
n_c^{\mathrm{lin}}
\simeq
\frac{
E^2\dfrac{1-r_b}{1+r_b}
}{
\Delta_a^2+
\kappa_a^2
\left(\dfrac{1-r_b}{1+r_b}\right)^2
}.
\label{eq:A9}
\end{equation}
Differentiating Eq.~(\ref{eq:A9}) with respect to $r_b$, we obtain
\begin{equation}
\frac{d n_c^{\mathrm{lin}}}{d r_b}
=-
\frac{2E^2}{(1+r_b)^2}
\frac{
\Delta_a^2-
\kappa_a^2
\left(\dfrac{1-r_b}{1+r_b}\right)^2
}{
\left[
\Delta_a^2+
\kappa_a^2
\left(\dfrac{1-r_b}{1+r_b}\right)^2
\right]^2
}.
\label{eq:A10}
\end{equation}
It follows that $d n_c^{\mathrm{lin}}/d r_b<0$ when
\begin{equation}
|\Delta_a|
>
\kappa_a\frac{1-r_b}{1+r_b}.
\label{eq:A11}
\end{equation}
For the parameters used in Fig.~2, $\Delta_a = \omega_b$ and $\kappa_a = 0.5\omega_b$. Since $0 \leq (1 - r_b)/(1 + r_b) \leq 1$ for $0 \leq r_b < 1$, the right-hand side of Eq.~(\ref{eq:A11}) does not exceed $0.5\omega_b$. Therefore, the condition is satisfied over the entire range of $r_b$ considered here.
After the transient evolution, the cavity field approaches a periodic steady state. The
quantity plotted in Fig.~\ref{fig2}(b) is
\begin{equation}
\overline{n}_c
=
\frac{1}{\tau}
\int_{t_0}^{t_0+\tau}|\alpha(t)|^2dt,
\label{eq:A12}
\end{equation}
where $\tau$ is the common modulation period. The numerical result is obtained from the full mean-field equations in Eqs.~(\ref{eq:A2}), with the OPA, MPA, and optomechanical terms retained. Fig.~\ref{fig2}(b) confirms that these additional periodic terms do not reverse the decreasing trend predicted by the linear-cavity estimate in Eq.~(\ref{eq:A10}). Therefore, in the parameter regime considered here, the entanglement
enhancement with increasing $r_b$ cannot be
attributed to an increase in the period-averaged intracavity photon number.

\section{\label{AppendixC}Entanglement enhancement at matched period-averaged intracavity photon number}

In this appendix, we describe the photon-number-matching procedure used in Fig.~\ref{fig4}(a) and decompose the feedback dependence of the maximum logarithmic negativity shown in Fig.~\ref{fig4}(b) in terms of the symplectic invariants of the CM. Both analyses are performed at $\theta=0$.

\subsection{Photon-number-matched reference system}

For every value of $r_b$, we construct a reference system without feedback by setting $r_b=0$ and adjusting its driving amplitude to $E_{\mathrm{match}}$. The matching condition is
\begin{equation}
\overline{n}_c^{(0)}
\left[E_{\mathrm{match}(r_b)}\right]
=
\overline{n}_c^{\mathrm{fb}}(r_b).
\label{eq:B2}
\end{equation}

All other system parameters, including the strengths, frequencies, and phases of the optical and mechanical parametric drives, are kept unchanged. The matched driving amplitude $E_{\mathrm{match}}$ is determined numerically from Eq.~\eqref{eq:B2}. For each matched pair, the periodic covariance matrices are calculated independently using the procedure described in Appendix~\ref{AppendixA}. 
The comparison therefore matches the period-averaged intracavity photon number between the two systems.

\subsection{Symplectic-invariant decomposition}

The CM of the COM system has the block form given in Eq.~(\ref{eq10}).
Then we introduce the symplectic invariants $I_1=\det V_a, I_2=\det V_b, I_3=\det V_{ab}, I_4=\det V$,
and $\widetilde{\Delta}=I_1+I_2-2I_3.$

Since the covariance matrix is periodic, its symplectic invariants depend on the time within one period. We introduce the normalized time
\begin{equation}
\xi=\frac{t-t_0}{\tau},
\qquad
0\leq \xi<1,
\label{eq:B4}
\end{equation}
where $t_0$ is chosen at the same reference phase of the parametric modulations for all values of $r_b$, and $\tau$ is the common period of the system.
At a given $\xi$ and $r_b$, the square of the smallest symplectic eigenvalue of the partially transposed covariance matrix is
\begin{equation}
\eta_-^2(\xi,r_b)
=
\frac{
\widetilde{\Delta}(\xi,r_b)
-
\sqrt{
\widetilde{\Delta}^2(\xi,r_b)
-
4I_4(\xi,r_b)
}
}{2}.
\label{eq:B5}
\end{equation}
For each value of $r_b$, we define
\begin{equation}
\xi_*(r_b)
=
\underset{0\leq \xi<1}{\operatorname{arg\,min}}\,
\eta_-(\xi,r_b),
\qquad
\eta_{-,\min}(r_b)
=
\eta_-[\xi_*(r_b),r_b].
\label{eq:B6}
\end{equation}
In the entangled region, the maximum logarithmic negativity over one period is therefore
\begin{equation}
E_{N,\max}(r_b)
=
-\ln\left[2\eta_{-,\min}(r_b)\right].
\label{eq:B7}
\end{equation}

Although the maximizing phase $\xi_*(r_b)$ generally varies with $r_b$, it does not contribute explicitly to the derivative of $E_{N,\max}$. For an isolated smooth maximum,
\begin{equation}
\left.
\frac{\partial E_N(\xi,r_b)}{\partial\xi}
\right|_{\xi=\xi_*}
=0.
\label{eq:B8}
\end{equation}
The chain rule therefore gives
\begin{equation}
\frac{dE_{N,\max}}{dr_b}
=
\left.
\frac{\partial E_N(\xi,r_b)}{\partial r_b}
\right|_{\xi=\xi_*}.
\label{eq:B9}
\end{equation}

Defining $R(\xi,r_b)=\sqrt{\widetilde{\Delta}^{\,2}(\xi,r_b)-4I_4(\xi,r_b)}$ and evaluating all quantities at $\xi=\xi_*(r_b)$, we obtain
\begin{equation}
\frac{dE_{N,\max}}{dr_b}
=
\Gamma_{\widetilde{\Delta}}
+
\Gamma_{I_4},
\label{eq:B11}
\end{equation}
with
\begin{equation}
\Gamma_{\widetilde{\Delta}}
=
\frac{1}{2R}
\left.
\frac{\partial\widetilde{\Delta}(\xi,r_b)}
{\partial r_b}
\right|_{\xi=\xi_*},
\label{eq:B12}
\end{equation}
and
\begin{equation}
\Gamma_{I_4}
=
-\frac{1}{2\eta_{-,\min}^{2}R}
\left.
\frac{\partial I_4(\xi,r_b)}
{\partial r_b}
\right|_{\xi=\xi_*}.
\label{eq:B13}
\end{equation}

The derivatives in Eqs.~\eqref{eq:B12} and \eqref{eq:B13} are evaluated at fixed $\xi$ and subsequently taken at $\xi=\xi_*(r_b)$. This procedure prevents small numerical shifts of the maximizing phase from introducing spurious discontinuities into the decomposition.

Here, $\Gamma_{\widetilde{\Delta}}$ describes the contribution from the feedback-induced variation of $\widetilde{\Delta}$, which combines the local covariance blocks and their cross correlations, whereas $\Gamma_{I_4}$ describes the contribution from $I_4=\det V$, which determines the global Gaussian purity. As shown in Fig.~\ref{fig4}(b), $\Gamma_{I_4}>0$ outweighs $\Gamma_{\widetilde{\Delta}}<0$ at small $r_b$. At larger $r_b$, $\Gamma_{I_4}$ becomes negative and eventually outweighs the positive $\Gamma_{\widetilde{\Delta}}$, producing the nonmonotonic behavior of $E_{N,\max}$. The agreement between $\Gamma_{\widetilde{\Delta}}+\Gamma_{I_4}$ and the direct numerical derivative provides a consistency check. Together with the photon-number-matched comparison in Fig.~\ref{fig4}(a), this result shows that the entanglement enhancement is not a trivial consequence of an increased period-averaged mean-field intracavity photon number, but reflects the feedback-modified covariance dynamics.

\section{\label{AppendixD}Covariance matrix analysis of directional steering}

In this appendix, we express the feedback-controlled transition from one-way to two-way steering shown in Fig.~\ref{fig6} in terms of the conditional covariance matrices. Using the block form of $V$ introduced in the main text, we define the two Schur complements as
\begin{equation}
\begin{aligned}
V_{b|a}
&=
V_b-V_{ab}^{T}V_a^{-1}V_{ab},\\
V_{a|b}
&=
V_a-V_{ab}V_b^{-1}V_{ab}^{T}.
\end{aligned}
\label{eq:conditional-covariances}
\end{equation}
Here, $V_{b|a}$ characterizes the conditional covariance of the phonon mode inferred from Gaussian measurements on the photon mode, whereas $V_{a|b}$ describes the reverse direction. The block-determinant identities give
\begin{equation}
\det V_{b|a}
=
\frac{\det V}{\det V_a},
\qquad
\det V_{a|b}
=
\frac{\det V}{\det V_b}.
\label{eq:conditional-determinants}
\end{equation}

Because the conditioned subsystem contains a single mode in either direction, its only symplectic eigenvalue is the square root of the corresponding determinant. Under the quadrature normalization used in this work, Gaussian steering from mode $j$ to mode $k$ occurs when this conditional symplectic eigenvalue is smaller than $1/2$. We therefore define
\begin{equation}
\begin{aligned}
\mathcal{E}_{a\rightarrow b}(t)
&=
2\sqrt{\det V_{b|a}(t)}
=
2\sqrt{\frac{\det V(t)}{\det V_a(t)}},\\
\mathcal{E}_{b\rightarrow a}(t)
&=
2\sqrt{\det V_{a|b}(t)}
=
2\sqrt{\frac{\det V(t)}{\det V_b(t)}}.
\end{aligned}
\label{eq:conditional-noise}
\end{equation}
Thus, $\mathcal{E}_{j\rightarrow k}(t)<1$ indicates steering from mode $j$ to mode $k$, while $\mathcal{E}_{j\rightarrow k}(t)=1$ defines the steering threshold. The quantities shown in Fig.~6 are related to the Gaussian-steering measures defined in the main text through
\begin{equation}
\begin{aligned}
\mathcal{G}_{j\rightarrow k}(t)
&=
\max\!\left\{0,-\ln\mathcal{E}_{j\rightarrow k}(t)\right\},\\
\mathcal{E}_{j\rightarrow k,\min}
&=
\min_{t\in[t_0,t_0+\tau)}
\mathcal{E}_{j\rightarrow k}(t),\\
\mathcal{G}_{j\rightarrow k,\max}
&=
\max\!\left\{0,-\ln\mathcal{E}_{j\rightarrow k,\min}\right\}.
\end{aligned}
\label{eq:periodic-steering-extrema}
\end{equation}

At $r_b=0$, $\mathcal{E}_{b\rightarrow a,\min}<1$, whereas $\mathcal{E}_{a\rightarrow b,\min}>1$, and hence only phonon-to-photon steering is present. As $r_b$ increases, $\mathcal{E}_{a\rightarrow b,\min}$ crosses unity at $r_b^*\simeq0.267$, while $\mathcal{E}_{b\rightarrow a,\min}$ remains below unity over the parameter range shown. Thus, $r_b^*$ marks the transition from one-way to two-way steering. Direct evaluation of the time-dependent conditional noises further confirms that, for $r_b>r_b^*$, there are times within each modulation period at which $\mathcal{E}_{a\rightarrow b}(t)<1$ and $\mathcal{E}_{b\rightarrow a}(t)<1$ simultaneously. The two steering directions therefore coexist rather than occurring at separate phases of the modulation.

\section{\label{Appendix E} Validity of the instantaneous-feedback approximation}

The analysis in the main text neglects the finite propagation time in the coherent feedback loop. To assess the validity of this approximation, we restore the loop delay $t_d$. The feedback phase $\theta$ defined in Sec.~\ref{sec:level2} includes the phase accumulated by the optical carrier during propagation, together with any static phase shifts introduced by the optical elements. Consequently, a fluctuation component at angular frequency $\nu$ relative to the carrier acquires the additional frequency-dependent phase factor $e^{i\nu t_d}$. The quantity $\nu t_d$ is therefore a delay-induced correction rather than an additional controllable phase.

Let
\begin{equation}
u(t)\equiv
\delta\widetilde{a}_{\mathrm{fb}}^{\mathrm{in}}(t)
\label{eq:delay-u}
\end{equation}
denote the optical fluctuation field incident on the cavity. Following the coherent-feedback construction and the standard input-output formalism~\cite{Lloyd2000,Agarwal2013}, the boundary conditions are
\begin{equation}
u(t)
=
t_b a^{\mathrm{in}}(t)
+
r_b e^{\mathrm{i}\theta}\delta a_{\mathrm{out}}(t-t_d),
\label{eq:delay-boundary}
\end{equation}
and
\begin{equation}
\delta a_{\mathrm{out}}(t-t_d)
=
\sqrt{2\kappa_a}\,\delta a(t-t_d)
-
u(t-t_d).
\label{eq:delay-io}
\end{equation}
Combining Eqs.~\eqref{eq:delay-boundary} and \eqref{eq:delay-io} gives the exact self-consistent delayed boundary condition
\begin{align}
u(t)+r_b e^{\mathrm{i}\theta}u(t-t_d)=t_b a^{\mathrm{in}}(t)
+\sqrt{2\kappa_a}\,r_b e^{\mathrm{i}\theta}\delta a(t-t_d).
\label{eq1}
\end{align}
Thus, the delay affects both the intracavity fluctuation and the field recirculating in the feedback loop.
Using the Fourier-transform convention
\begin{equation*}
o(t)
=
\int_{-\infty}^{\infty}
\frac{d\nu}{2\pi}\,
o[\nu]e^{-\mathrm{i}\nu t},
\end{equation*}
for which $o(t-t_d)\leftrightarrow e^{i\nu t_d}o[\nu]$, Eq.~\eqref{eq1} becomes
\begin{align}
u[\nu]
={}&
\frac{
\sqrt{2\kappa_a}\,r_b e^{\mathrm{i}\theta}e^{\mathrm{i}\nu t_d}
}{
1+r_b e^{\mathrm{i}\theta}e^{\mathrm{i}\nu t_d}
}
\delta a[\nu]+
\frac{
t_b
}{
1+r_b e^{\mathrm{i}\theta}e^{\mathrm{i}\nu t_d}
}
a^{\mathrm{in}}[\nu].
\label{eq:delay-frequency}
\end{align}
Equation~\eqref{eq:delay-frequency} retains the repeated circulation of both the cavity output field and the associated input noise.

When $|\nu t_d|\ll1$ over the relevant fluctuation bandwidth, one may set $e^{i\nu t_d}\simeq1$ while retaining the carrier-phase contribution contained in $\theta$. Transforming Eq.~\eqref{eq:delay-frequency} back to the time domain then gives
\begin{equation}
u(t)
=
\frac{
\sqrt{2\kappa_a}\,r_b e^{\mathrm{i}\theta}
}{
1+r_b e^{\mathrm{i}\theta}
}
\delta a(t)
+
\frac{
t_b
}{
1+r_b e^{\mathrm{i}\theta}
}
a^{\mathrm{in}}(t),
\label{eq:delay-instantaneous}
\end{equation}
which is the instantaneous-feedback relation used in the main text. Substitution of Eq.~\eqref{eq:delay-instantaneous} into the bare cavity Langevin equation recovers
\begin{equation}
\kappa_{\mathrm{fb}}
=
\kappa_a
\frac{1-r_b^2}
{1+r_b^2+2r_b\cos\theta},
\qquad
\Delta_{\mathrm{fb}}'(t)
=
\Delta_a'(t)
-
\frac{
2\kappa_a r_b\sin\theta
}{
1+r_b^2+2r_b\cos\theta
},
\label{eq:delay-effective-parameters}
\end{equation}
where
$\Delta_a'(t)=
\Delta_a-g[\langle b(t)\rangle+\langle b(t)\rangle^*]$.
Furthermore,
\begin{equation*}
\kappa_a
\left|
\frac{t_b}{1+r_b e^{i\theta}}
\right|^2
=
\kappa_{\mathrm{fb}},
\end{equation*}
showing that the same interference factor determines the effective cavity decay rate and the strength of the optical input noise.

At finite delay, the feedback self-energy and the external-noise transfer function entering the cavity equation are
\begin{equation}
\Sigma_{\mathrm{fb}}(\nu)
=
\frac{
2\kappa_a r_b e^{\mathrm{i}\theta}e^{\mathrm{i}\nu t_d}
}{
1+r_b e^{\mathrm{i}\theta}e^{\mathrm{i}\nu t_d}
},
\qquad
\mathcal{N}_{\mathrm{fb}}(\nu)
=
\frac{
t_b
}{
1+r_b e^{\mathrm{i}\theta}e^{\mathrm{i}\nu t_d}
}.
\label{eq:delay-transfer}
\end{equation}
Their expansions to first order in $\nu t_d$ are
\begin{subequations}
\begin{align}
\Sigma_{\mathrm{fb}}(\nu)
={}&
\frac{
2\kappa_a r_b e^{\mathrm{i}\theta}
}{
1+r_b e^{\mathrm{i}\theta}
}
+
\frac{
2i\kappa_a r_b e^{\mathrm{i}\theta}
}{
\left(1+r_b e^{\mathrm{i}\theta}\right)^2
}
\nu t_d
+
\mathcal{O}\!\left[(\nu t_d)^2\right],
\label{eq:delay-sigma-expansion}
\\
\mathcal{N}_{\mathrm{fb}}(\nu)
={}&
\frac{t_b}{1+r_b e^{\mathrm{i}\theta}}
\left[
1-
\frac{
i r_b e^{\mathrm{i}\theta}
}{
1+r_b e^{\mathrm{i}\theta}
}
\nu t_d
\right]
+
\mathcal{O}\!\left[(\nu t_d)^2\right].
\label{eq:delay-noise-expansion}
\end{align}
\label{eq:delay-expansions}
\end{subequations}

The term proportional to $i\nu$ in Eq.~\eqref{eq:delay-sigma-expansion} corresponds in the time domain to a correction
$-C_{\mathrm{fb}}\delta\dot{a}(t)$, where
\begin{equation*}
C_{\mathrm{fb}}
=
\frac{
2\kappa_a r_b e^{\mathrm{i}\theta}t_d
}{
\left(1+r_b e^{\mathrm{i}\theta}\right)^2
}.
\end{equation*}
Let $\Omega_{\mathrm{rel}}$ denote the largest angular-frequency scale that contributes appreciably to the slowly varying fluctuation dynamics. The relevant dimensionless delay parameters may then be written as
\begin{equation}
\epsilon_{\mathrm{ph}}
=
\Omega_{\mathrm{rel}}t_d,
\qquad
\epsilon_{\mathrm{in}}
=
\frac{
r_b\Omega_{\mathrm{rel}}t_d
}{
\left|1+r_b e^{\mathrm{i}\theta}\right|
},
\qquad
\epsilon_{\mathrm{dyn}}
=
\frac{
2\kappa_a r_b t_d
}{
\left|1+r_b e^{\mathrm{i}\theta}\right|^2
}.
\label{eq:delay-small-parameters}
\end{equation}
Here, $\epsilon_{\mathrm{ph}}$ controls the expansion of the delay phase, $\epsilon_{\mathrm{in}}$ bounds the relative correction to the optical noise-transfer function, and $\epsilon_{\mathrm{dyn}}=|C_{\mathrm{fb}}|$ measures the delay-induced correction to the cavity time derivative. A sufficient condition for the instantaneous-feedback approximation is therefore
\begin{equation}
\max
\left\{
\epsilon_{\mathrm{ph}},
\epsilon_{\mathrm{in}},
\epsilon_{\mathrm{dyn}}
\right\}
\ll1.
\label{eq:delay-criterion}
\end{equation}

We now evaluate Eq.~\eqref{eq:delay-criterion} for the parameter range considered in the main text. The numerical calculations use
$\omega_b/2\pi=1~\mathrm{MHz}$,
$\kappa_a/2\pi=0.5~\mathrm{MHz}$,
and $\Delta_c=1.18\omega_b$, with
$\omega_m/\Delta_c\leq2$.
The largest explicit modulation frequency is therefore
$2\Delta_c=2.36\omega_b$. Since the parametric gains and decay rates are smaller than this scale, we take
$\Omega_{\mathrm{rel}}=2.36\omega_b$
as the characteristic bandwidth of the dominant fluctuation components.

For the total feedback-loop optical path length
$L=10~\mathrm{cm}$ considered in Sec.~\ref{sec:level4}, the propagation delay is
\begin{equation*}
t_d=\frac{L}{c}
\simeq3.34\times10^{-10}~\mathrm{s}
=0.334~\mathrm{ns}.
\end{equation*}
Over all stable parameter points used in the numerical calculations, $r_b\leq0.70$. The phase-independent bound
\begin{equation*}
\left|1+r_b e^{\mathrm{i}\theta}\right|
\geq1-r_b
\geq0.30
\end{equation*}
then gives
\begin{equation}
\epsilon_{\mathrm{ph}}
\lesssim4.95\times10^{-3},
\qquad
\epsilon_{\mathrm{in}}
\lesssim1.16\times10^{-2},
\qquad
\epsilon_{\mathrm{dyn}}
\lesssim1.64\times10^{-2}.
\label{eq:delay-numerical}
\end{equation}
Thus, even under this conservative phase-independent estimate, the largest delay correction is less than $1.7\times10^{-2}$. The parameter range and feedback-loop length considered in this work therefore satisfy Eq.~\eqref{eq:delay-criterion}, validating the instantaneous-feedback approximation used in the mean-field equations and in the fluctuation dynamics.

\bibliography{1}

\end{document}